\input epsf.tex
\documentstyle[twocolumn,prb,aps,floats]{revtex}
\begin{document}
\draft
\title{Elastic Study of Antiferromagnetic Fluctuations in the Layered
Organic Superconductors $\kappa$-(BEDT-TTF)$_2$X}
\author{K. Frikach, M. Poirier, M. Castonguay and K.D. Truong}

\address{Centre de Recherche sur les Propri\'et\'es \'Electroniques de
Mat\'eriaux Avanc\'es and D\'epartement de Physique,}
\address{Universit\'e de Sherbrooke, Sherbrooke, Qu\'ebec, Canada J1K 2R1.}

\date{\today}
\maketitle
\begin{abstract}

In an ultrasonic experiment, we have investigated the temperature
profile of the velocity of longitudinal elastic waves propagating along
a direction perpendicular to the layers in the organic superconductors
$\kappa$-(BEDT-TTF)$_2$X, X = Cu(SCN)$_2$ and Cu[N(CN)$_2$]Br. Although
a small decrease of the velocity is observed at the superconducting
transition, the most anomalous behavior is obtained in the normal
metallic state where an important softening  is identified around 
40-50 K. In order to characterize the origin of this anomaly, we have studied its
behavior under the application of hydrostatic pressure. The observed
behavior is found to mimic those of the transport and magnetic properties of
these materials which have been attributed to the magnetic
fluctuations. Following the example of one-dimensional insulating
systems where coupling between longitudinal acoustic waves and magnetic
fluctuations is known to occur, our results suggest that the pseudo-gap
regime of these two-dimensional organic superconductors is dominated by
a similar mechanism.

\end{abstract}

\pacs{74.70.Kn, 74.25.Ld, 74.40.+k, 74.62.Fj}

Organic compounds of the $\kappa$-(BEDT-TTF)$_2$X family are
highly anisotropic, layered, extreme type II superconductors which
show the highest superconducting transition temperatures known to
date.\cite{Williams}  They are continuing to attrack considerable
attention because of their similarity to the high T$_c$ cuprates and the
possibility that they also have a non-conventional pairing
state.\cite{McKenzie} In these materials the superconducting phase is in
close proximity to an antiferromagnetic (AF) phase. Indeed, at ambient
pressure, the compounds with X = Cu[N(CN)$_2$]Br and Cu(SCN)$_2$ are
showing superconducting ground states around 10 K, while an insulating
one with antiferromagnetic ordering is obtained below 25 K for X =
Cu[N(CN)$_2$]Cl.\cite{Urayama,Welp,Lefebvre}  However, the latter compound shows
also a superconducting ground state around 12 K under a small
hydrostatic pressure of 300 bar.\cite{Williams2} In view of the strong
dimerization of the BEDT-TTF molecules, this pressure induced metal-insulator
transition results from the competition between the repulsive
Coulomb energy (U) and the bandwidth (W) in the effectively half-filled
electronic band structure (Mott transition).\cite{Kanoda} In the
Cu[N(CN)$_2$]Br and Cu(SCN)$_2$ salts, the proximity of the AF magnetic phase can
be infered from the transport and magnetic properties of the normal
state.  In NMR experiments the maximum of (T$_1$T)$^{-1}$ around 50
K and the rapid decrease of the Knight shift below 50 K were attributed
to both AF magnetic fluctuations and to a pseudo-gap.\cite{Kawamoto,Mayaffre}
At the same temperature, the intra- and inter-layer resistivity data
suggest a change of regime (peak in dR/dT) around the same
temperature.\cite{Shushko} All these features observed on the magnetic
and transport properties are highly sensitive to pressure and they
completely disappear above 2 kbar.\cite{Mayaffre,Shushko}

Ultrasonic techniques can be used to study quasiparticle and magnetic
excitations effects on the elastic properties of superconductors and
attenuation measurements are probably the key experiments that are still to
be performed on organic and high T$_c$ superconductors to assess the
nature of the superconducting state. Indeed, the smallness of the
crystals prohibits the application of standard ultrasonic methods to
these materials. Nevertheless few measurements of the ultrasonic
velocity\cite{Yoshizawa,Frikach} in the superconducting state have been
performed on organic materials. In this paper a modified ultrasonic
technique is used to investigate the elastic velocity in organic
superconductors $\kappa$-(BEDT-TTF)$_2$X, X = Cu(SCN)$_2$ and
Cu[N(CN)$_2$]Br in their normal state. In addition to a small velocity
anomaly found at the superconducting transition temperature, we
observe, for both compounds, a very large softening in the normal
state. This peak is likely the result of a coupling between magnetic
fluctuations and acoustic phonons. An investigation of these anomalies
under hydrostatic pressure allows to obtain a phase diagram which
establishes a clear connection between the temperatures of the
superconducting transition and the pseudo-gap.

Single crystals of the organic superconductors $\kappa$-(BEDT-
TTF)$_2$Cu[N(CN)$_2$]Br and $\kappa$-(BEDT-TTF)$_2$Cu(SCN)$_2$ were
synthesized by the electrocrystallization technique described
elsewhere.\cite{Kini}  Because of their quasi-two-dimensional
structure, the crystals have generally the shape of small shiny
platelets or cubes.  In our ultrasonic technique, the parallel faces
identifying the highly conducting planes are the only faces that can
be used properly to generate and detect acoustic waves propagating along
a direction perpendicular to the layers. The crystals with typical
surface area of 1 mm$^2$ (layer plane) and thickness of 0.3 mm (transverse
direction) were used in a pulse transmission experiment. The usual pulse
echo method cannot, however, be utilized directly since the
propagation length ($\sim$ 0.3 mm) does not allow time separation of
transmitted and reflected echoes. We have thus used a modified set-up
for the measurement. One parallel face of the crystal is first glued on the surface of a
CaF$_2$ delay line (buffer length $\sim$ 7 mm).  Then, a LiNbO$_3$ piezoelectric
transducer bonded on the other parallel face generates longitudinal
waves, at 30 MHz and odd overtones, that will propagate through the
crystal-delay line ensemble and will be detected by a second piezoelectric
transducer at the other end of the buffer. The ultrasonic velocity is
measured with a pulsed acoustic interferometer.\cite{Poirier} GE silicon
sealant was used for all the bonds.  The ultrasonic experiment can also
be performed in a pressure cell using a liquid (maximum value 8
kbar). The temperature is monitored with a Si diode sensor and stabilized with a
LakeShore controller. Magnetic field measurements up to 9 Tesla can also
be done with this set-up.

\input epsf.tex \vglue 0.4 cm
\epsfxsize 8 cm
\begin{figure}
\centerline{\epsfbox{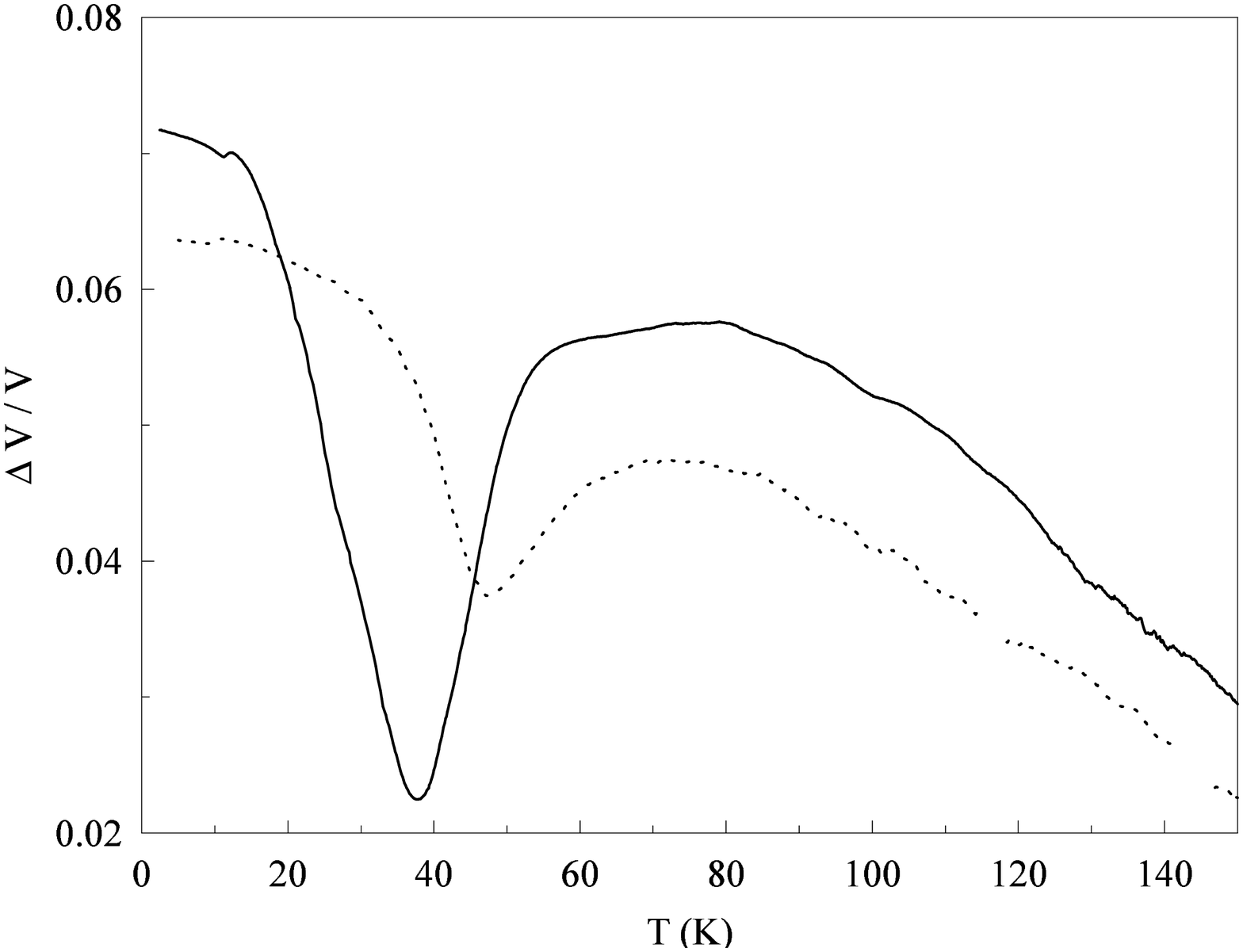}}
\caption{Relative variation of the longitudinal velocity propagating
along a direction perpendicular to the layers as a function of
temperature: $\kappa$-(BEDT-
TTF)$_2$Cu[N(CN)$_2$]Br (full line) and $\kappa$-(BEDT-
TTF)$_2$Cu(SCN)$_2$ (dotted line).}
\label{}
\end{figure}

The velocity of longitudinal waves propagating along a direction
perpendicular to the layers has a rather low value around 2000 m/sec, in
agreement with the 2D character of the structure. In this work we are
interested in the temperature profile of the velocity only.  This is obtained
by monitoring the phase ($\phi$) of the transmitted signal as a function of the
temperature. However, since two terms,  $\phi = k_1 l_1 + k_2 l_2$ where
$k_1$, $k_2$, $l_1$ and $l_2$ are respectively the ultrasonic wave
vectors and the lengths of the crystal (1) and the delay line (2),
contribute to the total phase of the transmitted signal, it is
necessary to substract the delay line contribution ($k_2 l_2$) to isolate the
crystal one ($k_1 l_1$).  The latter is measured in a pulsed reflection experiment
in the delay line only at the same frequency and  for identical experimental
conditions.  The crystal velocity data have not been corrected to take into account
thermal expansion effects since these are orders of magnitude too
small.\cite{Kund,Lang} In the range 30-210 MHz, no frequency effects could be
detected on the ultrasonic velocity data. All the data presented in this
paper have been obtained at 100 MHz.

We present in figure 1 the relative change of the velocity, $\Delta V /
V = (V - V_0) / V_0$ where $V_0$ is the velocity at 200 K, as a function
of temperature in the range 2-150 K for the $\kappa$-(BEDT-
TTF)$_2$Cu[N(CN)$_2$]Br and $\kappa$-(BEDT-TTF)$_2$Cu(SCN)$_2$ organic
superconductors obtained at a pressure of a few mbar of helium gas.  For
both compounds, a weak softening of the velocity (10$^{-3}$) is observed at the
superconducting transition temperature T$_c$ = 11.3 and 9.2 K respectively for
the Cu[N(CN)$_2$]Br and Cu(SCN)$_2$ crystals.  A similar type of anomaly
in the superconducting state has previously been reported in both
compounds.\cite{Yoshizawa,Frikach} As the temperature is increased above
T$_c$, both velocity profiles reveal a very $\it large$ dip centered at 38
and 50 K respectively for the Cu[N(CN)$_2$]Br and Cu(SCN)$_2$ compounds.
The relative velocity softening is much larger for the Cu[N(CN)$_2$]Br
crystal ($\sim$ 6\%) than for the Cu(SCN)$_2$ one ($\sim$ 2\%). No
additional anomalous behavior is observed up to room temperature where
the profile is dominated by the usual anharmonic contribution. As
observed in transport experiments, rapid cooling of the crystals below
90 K decreases slightly the superconducting temperature and shifts a
little the dip to higher temperatures. Although the application of a
magnetic field perpendicularly to the layers decreases T$_c$ and
eventually suppresses completely the superconducting state, no effects
could be detected on the softening anomaly up to 9 Tesla . It is worth
mentioning that recent measurements of thermal expansion\cite{Lang} have
also revealed an anomalous behavior of that quantity along the direction
perpendicular to the layers in the Cu[N(CN)$_2$]Br crystal: a peaked
anomaly centered around 40 K was observed in agreement with our results.

A few scenarios can be suggested to explain the anomalous elastic
behavior of these superconductors in the normal state. First, this could
be the result of a structural phase transition.  However, no such
transition has ever been reported in either of these compounds,
although indications of a second-order transition around 80 K have been
obtained in several experiments\cite{Kund,Su} in relation to the onset of
conformational order among the terminal ethylene groups of the donor
molecules. The appearance of a structurally disordered state below 80 K
can be induced easily by a rapid cooling process. This has been also verified
on our ultrasonic velocity data as mentioned previously. The data
presented here were obtained with the slowest possible cooling in order
to avoid disorder. Although structural order clearly affects the low
temperature properties, they cannot be responsible for the observed
elastic feature around 40-50 K. 

In a metal, the ultrasonic wave can also interact with the conduction
electrons: the softening  or hardening of an elastic constant (and the
velocity) is then related to the augmentation or reduction in the
quasiparticle screening of ion potential. In order to appreciate
possible quasiparticle screening in our crystals, we present in figure 2
the microwave transverse resistivity of both compounds as a function of
temperature. Even if the absolute values are 2 to 3 orders of magnitude
higher, the temperature profile shown here is identical to the in-plane
resistivity one;\cite{Shushko} this gives an indication that the transverse transport
is diffusive in these highly anisotropic conductors. As the temperature
is decreased below the resistivity maximum around 80-90 K, a metallic
regime first sets in and the resistivity decreases with a rate dR/dT which is
the largest at respectively 38 and 50 K for the Cu[N(CN)]$_2$Br and
Cu(SCN)$_2$ crystals. Then, the expected decrease of the resistivity is
observed when entering in the superconducting state at T$_c$. Since the
softening elastic dip and the resistivity decrease in the normal state
occur exactly at the same temperature values in both compounds, this is
suggestive of a common mechanism.  However no direct correlation can be
established between these two features.  Indeed, quasiparticle screening cannot
explain an elastic softening which shows the temperature profile of a
dip and, moreover, it should yield a larger acoustic anomaly when the compounds
enter in the superconducting state, an observation which is not supported by
the data presented in Fig.1. 

\input epsf.tex \vglue 0.4 cm
\epsfxsize 7 cm
\begin{figure}
\centerline{\epsfbox{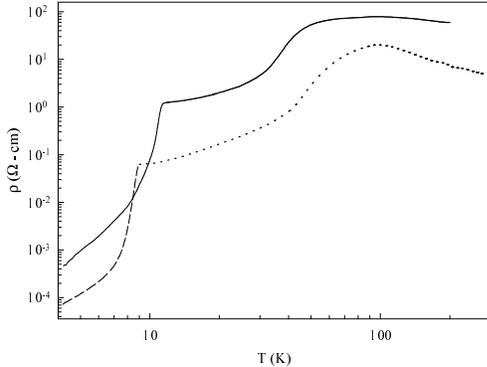}}
\caption{Microwave transverse resistivity (17 GHz) as a function of
temperature: $\kappa$-(BEDT-
TTF)$_2$Cu[N(CN)$_2$]Br (full line) and $\kappa$-(BEDT-
TTF)$_2$Cu(SCN)$_2$ (dotted line).}
\label{}
\end{figure}

In transport measurements,\cite{Shushko} it has been suggested that
the rapid decrease of the resistivity around 40-50 K is related to the
presence of both a pseudo-gap regime and AF magnetic fluctuations in the
normal state. These fluctuations
have been clearly identified in NMR experiments.\cite{Kawamoto,Mayaffre}
In one-dimensional (1D) insulating systems,\cite{Trudeau,Dumoulin} it is
known that AF magnetic fluctuations can couple to longitudinal acoustic
phonons and yield a softening of the ultrasonic longitudinal velocity along the
1D direction. The magneto-elastic coupling is then the result of the
modulation of the spin exchange constant $J$ by the acoustic phonons.  In
such a 1D model, the softening occurs within a rather wide temperature
domain with a maximum at $T \sim J$. Now one can ask the following question: is it
possible to support a similar explanation in a 2D conducting system in which
AF magnetic fluctuations are present? We will try to answer this
question by investigating further the softening observed on the
ultrasonic velocity by applying hydrostatic pressure. From NMR
experiments (nuclear relaxation and Knight shift), the magnetic
fluctuations and the pseudo-gap in these organic superconductors are indeed suppressed by a
pressure of a few kbar.\cite{Mayaffre,Shushko}  We present in figure 3
the temperature profile of the ultrasonic velocity at different
pressures in the range 0.4-2.5 kbar. The observed trend with pressure is
identical for both compounds: in addition to the decrease of the
superconducting transition temperature T$_c$ with increasing pressure, there
is a depression and a shift of the dip to higher temperatures
in the normal state. No anomaly can be detected above 2.5 kbar,
although the superconducting transition is still observed.  These data
confirm that the elastic anomaly shows temperature and pressure
dependences which are similar to the ones observed on the transport
and magnetic properties. We thus believe that it can be the result of a
coupling between acoustic phonons and magnetic fluctuations. The absence
of magnetic field effects (up to 9 Tesla) on the anomaly is not
necessarily in contradiction with the magneto-elastic coupling picture.
Indeed, field effects would be expected to be observed only if the Zeeman
field energy is approaching $J_{\perp}$, namely the interlayer exchange
constant which is not known for the moment.

\input epsf.tex \vglue 0.4 cm
\epsfxsize 7 cm
\begin{figure}
\centerline{\epsfbox{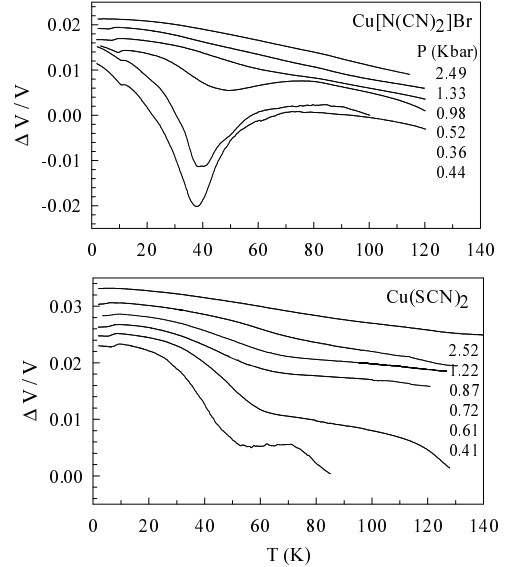}}
\caption{Relative variation of the longitudinal velocity propagating
along the direction perpendicular to the layers as a function of
temperature at different pressure values.}
\label{}
\end{figure}

\input epsf.tex \vglue 0.4 cm
\epsfxsize 7 cm
\begin{figure}
\centerline{\epsfbox{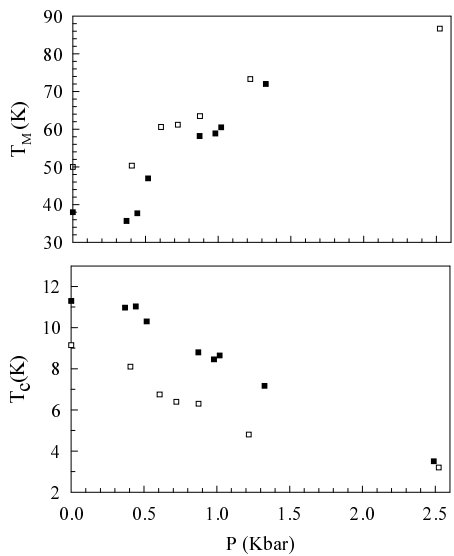}}
\caption{Phase diagram of $\kappa$-(BEDT-
TTF)$_2$Cu[N(CN)$_2$]Br (black squares) and $\kappa$-(BEDT-
TTF)$_2$Cu(SCN)$_2$ (white squares) organic conductors: superconducting
transition temperature T$_c$ (lower panel) and softening peak
temperature T$_M$ (upper panel).}
\label{}
\end{figure}

In figure 4 we show the phase diagram which is obtained for the two
organic conductors when the superconducting temperature T$_c$
(lower panel) and the temperature of the dip T$_M$ (upper
panel) are plotted as a function of pressure.  The variation of T$_c$
with pressure is in full agreement with already published
data.\cite{Schriber,Sushko2} The variation of +25 K/kbar for T$_M$ with
pressure is similar for the two materials. Although there is a 12
K separation between the T$_M$'s at zero pressure, this difference tends
to be reduced as pressure is increased; a similar trend is observed for
the T$_c$'s.  This diagram is coherent with the fact that, in these
materials, the superconducting phase is in close proximity to an
antiferromagnetic phase. Moreover, magnetic fluctuations appear to be an
important ingredient to obtain a superconducting state at high
temperatures. Indeed, the Cu[N(CN)]$_2$Br crystal presents a higher
T$_c$ than the Cu(SCN)$_2$ one at zero pressure, while the magnetic
fluctuation effects manifest themselves at a lower temperature T$_M$.
Finally, it is worth mentioning a special feature observed for the
Cu[N(CN)]$_2$Br crystal. In figure 3, we notice that the anomaly
obtained at 0.44 kbar has a larger amplitude and a smaller T$_M$ than
the 0.36 kbar one. For the moment we cannot decide if this effect is
real or not because we do not have enough precision on the pressure; one
will need a gas pressure cell instead of a liquid to overcome
this difficulty.

In summary, we have identified an important softening  on the
temperature profile of longitudinal ultrasonic waves propagating along a
direction perpendicular to the conducting layers in 2D organic superconductors. Its
pressure dependence suggests that this anomaly is related to the presence
of magnetic fluctuations in this temperature range, as it was previously
observed on the magnetic properties.  The anomaly could
then result from a coupling between acoustic phonons and AF magnetic
fluctuations. Although similar anomalies have been observed in 1D
insulating magnetic systems, our data give the first observation of such
a phenomenon in an organic conductor having a 2D character. Considering
the large amplitude of the anomaly, this signifies an important coupling
between the lattice and the spin degrees of freedom.

The authors thank C. Bourbonnais for useful discussions and critical
reading of the manuscript and Guy Quirion for his expertise relative to
the pressure measurements. This work was supported by grants from the
Fonds pour la Formation de Chercheurs et l'Aide \`a la Recherche of the
Government of Qu\'ebec (FCAR) and from the Natural Science and
Engineering Research Council of Canada (NSERC).

\end{document}